\documentclass[onecolumn,aps,11pt,showpacs,floatfix,nofootinbib]{revtex4}
\usepackage{amssymb,amsmath,mathrsfs}
\usepackage{subfigure}
\usepackage{color}
\usepackage{ucs}
\usepackage[utf8x]{inputenc}
\usepackage{textcomp}
\usepackage[T1]{fontenc}
\usepackage{mathtools}

\newcommand{\doe}{\psi\ot\lam}

\newcommand{\dod}{\psi_1\ot\psi_2}

\newcommand{\II}{\mathbb{I}}

\newcommand{\n}{\nonumber\\}

\newcommand{\mpp}{\mathscr{P}}
\newcommand{\mcc}{\mathscr{C}}
\newcommand{\mttt}{\mathscr{T}}

\newcommand{\mot}{\mathcal{T}}

\newcommand{\sst}{\scriptscriptstyle}
\newcommand{\bec}{\begin{center}}
\newcommand{\eec}{\end{center}}

\newcommand{\bea}{\begin{array}}
\newcommand{\ear}{\end{array}}

\newcommand{\eee}{\mathcal{E}}\newcommand{\ddd}{\mathcal{D}}\newcommand{\sss}{\mathcal{S}}

\newcommand{\bfr}{\begin{flushright}}

\newcommand{\efr}{\end{flushright}}
\newcommand{\noi}{\noindent}

\newcommand{\pp}{\mathcal{P}}
\newcommand{\oo}{\mathcal{O}}
\newcommand{\cc}{\mathcal{C}}
\newcommand{\ttt}{\mathcal{T}}

\newcommand{\cl}{{\mt{C}}\ell}

\newcommand{\ot}{\otimes}

\newcommand{\bege}{\begin{equation}}
\newcommand{\enge}{\end{equation}}

\newcommand{\beq}{\begin{eqnarray}}\newcommand{\benu}{\begin{enumerate}}\newcommand{\enu}{\end{enumerate}}
\newcommand{\eeq}{\end{eqnarray}}
\newcommand{\mt}{\mathcal}

\newcommand{\ZZ}{\mathbb{Z}}

\newcommand{\lam}{\lambda}

\newcommand{\OO}{\mathbb{O}}

\newcommand{\bx}{\begin{pmatrix}}
\newcommand{\ex}{\end{pmatrix}}

\usepackage{amsfonts}

\begin{document}

\title{VSR symmetries in the DKP algebra: the interplay between Dirac and Elko spinor fields}
\author{R. T. Cavalcanti}
\email{rogerio.cavalcanti@ufabc.edu.br}
 \affiliation{
Centro de Ci\^encias Naturais e Humanas,
UFABC, 09210-580, Santo Andr\'e - SP, Brazil}
\author{J. M. Hoff da Silva}
\email{hoff@feg.unesp.br} \affiliation{UNESP - Campus de Guaratinguet\'a - DFQ, Av. Dr.
Ariberto Pereira da Cunha, 333, 12516-410, Guaratinguet\'a - SP,
Brazil}
\author{Rold\~ao da Rocha}
\email{roldao.rocha@ufabc.edu.br} \affiliation{
Centro de Matem\'atica, Computa\c c\~ao e Cogni\c c\~ao,
UFABC, 09210-180, Santo Andr\'e - SP, Brazil}
\affiliation{International School for Advanced Studies (SISSA), Via Bonomea 265, 34136 Trieste, Italy}

\pacs{04.20.Gz, 03.65.Fd, 11.10.-z}

\begin{abstract}
VSR symmetries are here naturally incorporated in the DKP algebra on the spin-0 and the spin-1 DKP sectors. 
We show that the Elko (dark) spinor fields structure plays an essential  role on accomplishing this aim, unravelling hidden symmetries on the bosonic DKP fields under the action of discrete symmetries. 
 
\end{abstract}
\maketitle

\section{Introduction}

Elko spinor fields (dual-helicity eigenspinors of
the charge conjugation operator \cite{allu,allu1}) are  spin-1/2  matter fields, with unexpected properties that make them prime candidates to describe dark matter. 
Recent efforts to scrutinize the underlying  structure of Elko fields, that incorporate both the Very Special Relativity (VSR) paradigm \cite{cohen} and dark matter as well, have been reported in \cite{horv1}.
In this context, an Elko spinor mass generation mechanism was proposed in  \cite{alex}, via a natural coupling to the kink solution of a $\lambda \phi^{4}$ field theory, regarding  exotic couplings between Elko spinor fields and scalar field topological solutions \cite{alex}.
Some attempts to detect Elko at the LHC have been moreover proposed \cite{marcao}, as well as promising  applications  \cite{exotic,saulo,sh,bur1,bur2}.

{Elko spinor fields occupy just one type among other classes of spinor fields that encompass the well known Dirac, Weyl, Majorana, {flag-poles}, dipoles, and flag-dipoles ones \cite{lou2,dipoles}. Elko has provided prominent applications {in} cosmology, gravity  \cite{saulo,sh,sadj,liu,jmp07,abla,qua,boeh,vign,fabb1}),} field theory \cite{alex,horv1,lee1,lee2,delag} and its further supersymmetric formulation {as well} \cite{wund}{. T}hose classes have been shown worth to further explore, and the phenomenology related to new fermionic fields   has been recently investigated  \cite{marcao,alve}. Moreover, in the framework  of  $f(R)$ and ESK gravity, the Dirac equation admits new solutions distinct from Dirac spinor fields, which are flag-dipoles spinor fields  \cite{desk,esk}. Black hole thermodynamics 
has been also investigated in the context of 
Elko spinor fields \cite{hawk1}.

The aim of this article is to evince hidden symmetries that underlie bosonic fields described by the spacetime DKP (Duffin-Kemmer-Petiau) algebra. This algebra is here a subalgebra of a bigger space formed  by the tensor product of two Clifford algebras that comprise both Dirac and Elko spinor fields. The unexpected behavior of DKP spinor fields  under discrete symmetries  show  that the full Lorentz group is not the symmetry group associated to the whole DKP algebra. Instead, merely the subgroups HOM(2) and SIM(2) are related to VSR symmetries\footnote{The  incorporation of the parity and the time reversal operators  widens the VSR  subgroups to the full Lorentz group.}. Rephrasing it, although Cohen and Glashow state that VSR implies special relativity either in the context of local
quantum field theory or in the framework of CP conservation \cite{cohen}, merging VSR with Elko is still feasible \cite{crevice}. The
amplitude for the two body decay of a spinless particle at
rest may depend
on the direction of the decay products relative to the VSR preferred
direction \cite{cohen,crevice}, and  VSR signature is expected to arise for the mass dimension one Fermi field~\cite{marcao}. While a new connection with the VSR has begun to be evinced \cite{horv1,crevice}, we stress that the HOM(2) is both sufficient and necessary to encompass  the negative outcome 
of the Michelson-Morley experiment \cite{cohen}.

This paper is organized as follows: in the next Section we review the basic features of the DKP algebra and present  it as the product of two Clifford algebras, in such a way that DKP fields can be presented as the tensor product between spinor fields. In Section III, Elko spinor fields are briefly revisited. The  types of spinor fields (under Lounesto spinor field classification) that are preserved under sum of spinors, are discussed in Section IV. Moreover, the 
conditions that the Elko and Dirac spinor fields must satisfy
for the correct definition of the DKP fields, in terms of the tensor product of spinor fields, are obtained. This is accomplished in order to assure that the spinor fields classes, under the Lounesto spinor field classification, are preserved. Elko and Dirac spinors are shown to be building blocks of the DKP algebra,  by revealing how the Elko unusual features reflects at the bosonic level of the symmetries of the DKP algebra. The DKP algebra can be obtained by the tensor product between Dirac spinor fields,  and thus preserves Lorentz symmetries.    The VSR symmetries, underlying the Elko spinor fields structure \cite{cohen,horv1}, are thus induced on the DKP algebra. In the last Section we conclude.
 
\section{The DKP Algebra}
\label{adkp}
The study of elementary particles by means of classical wave theory depends on the charged (either electric charge or moment dipolar charge) or uncharged aspect of the particle under consideration.  A particle of mass $m$ 
 is said to be a meson if it is described by the  wave equation in Minkowski spacetime
\beq
\left(\eta^{\mu\nu}\beta_\mu\partial_\mu + im\right)\psi=0,
\eeq
where  $\beta_\mu$ satisfy the
commutation rules proposed in \cite{duffin,duffin1,duffin2}:
\beq
\beta_\mu\beta_\nu\beta_\rho + \beta_\rho\beta_\nu\beta_\mu = \eta_{\nu\rho}\beta_\mu + \eta_{\nu\mu}\beta_\rho.\eeq\noi 
In addition, the Dirac equation is well known 
to be written as 
\beq
\left(\eta^{\mu\nu}\gamma_\mu\partial_\mu + im\right)\psi=0,
\eeq
denoting the set $\{\gamma _{\mu }\}$ of Dirac matrices in Eq.(\ref{dirac matrices}), and the set $\{\mathbf{1},\gamma _{\mu },\gamma _{\mu }\gamma _{\nu },\gamma _{\mu }\gamma _{\nu
}\gamma _{\rho },\gamma _{0}\gamma _{1}\gamma _{2}\gamma _{3}\}$ ($\mu ,\nu
,\rho =0,1,2,3$, and $\mu <\nu <\rho $) as  a basis for ${\rm M}(4,\mathbb{C})$, such that   $\gamma _{\mu }\gamma _{\nu }+\gamma _{\nu }\gamma
_{\mu }=2\eta _{\mu \nu }\mathbf{1}$. The Clifford product is denoted
by juxtaposition.

The Dirac equation is usually written in a form that can be immediately compared with the meson wave equation, by the prescription $\beta_\mu \mapsto \gamma_\mu$.
The massless DKP theory can not be obtained as a zero mass limit of the
massive DKP case. Some authors consider the Harish-Chandra Lagrangian density for
the massless DKP theory in Minkowski spacetime, given
by \cite{HCC}
\begin{equation}\label{lag}
\mathcal{L}=i\bar{\psi}\tau \beta ^{\mu}\partial _{\mu}\psi -i\partial _{\mu}%
\bar{\psi}\beta ^{\mu}\tau \psi -\bar{\psi}\tau \psi \;, 
\end{equation}%
where  $\tau $ is a {singular} idempotent matrix satisfying
\begin{equation}\label{gama} \beta ^{\mu}\tau +\tau \beta ^{\mu}=\beta ^{\mu},\qquad  {\rm and}\qquad\;\;\;  \tau ^{\dag }=\tau\,,\end{equation}\noindent and moreover ${\beta ^{0}}^{\dag }={\beta ^{0}}$, ${%
\beta ^{i}}^{\dag }=-{\beta ^{i}}$. 
Besides, the angular momentum is provided by  $
S^{\mu\nu}=\beta ^{\mu}\beta ^{\nu} - \beta^\nu\beta^\mu$ \cite{duffin,duffin1,duffin2}.

By denoting hereon a $n \times n$  matrix $B$ by $\mathbb{B}_n$, the spin-$0$ sector of the DKP algebra is realized when a specific representation of the DKP algebra is used, $%
\tau = \{0\}\oplus \mathbb{I}_4$. It thus  reproduces the massless Klein-Gordon-Fock field \cite{CQG}.  The scalar sector of
the massless DKP theory associated to the 5-dimensional representation of massless DKP algebra provides  
\begin{equation}\label{psid0}
\psi =(\varphi, A ^{\mu}) ^{\intercal},
\end{equation}%
where $\varphi $ and $A ^{\mu}$ are 
scalar and  4-vector under {Lorentz} transformations, respectively.

The spin-1 DKP field $\psi$ can be provided by choosing  
 $
\tau = \mathbb{O}_4\oplus\mathbb{I}_6$, therefore the DKP field  is
thus a 10-component column vector
\begin{equation}\label{psid1}
\psi =(
\psi ^{\mu}, F ^{\mu\nu}) ^{\intercal},
\end{equation}%
where $\psi ^{\mu}$ and $F ^{\mu\nu}$ are respectively a 4-vector and some  antisymmetric tensor in Minkowski spacetime.
In other words, the above Lagrangian yields to the DKP wave equation 
\begin{equation}
\eta^{\mu\nu}\beta_{\mu}\partial_{\nu}\psi-\tau\psi=0\;. \label{llag}%
\end{equation}
In order to settle on the spin 0 sector the so called  Umezawa's
projectors $P$ and $P^{\mu}$  \cite{Umezawa} are applied on
 Eq.(\ref{llag}), and taking into account the above relations (\ref{gama})
implying $\tau P=P\tau$ and $P^{\mu}\tau+\tau P^{\mu}=P^{\mu}$, the
equation of motion for the massless scalar field $P\psi$ reads
$\partial_{\mu}\partial^{\mu}\left(  P\psi\right)  =0.$ 
Usually, the gauge invariance is derived when  a representation for $\beta^{\mu}$ is employed in which
\begin{equation}
\tau={\rm diag}(\alpha,1-\alpha,1-\alpha,1-\alpha,1-\alpha).
\end{equation}
In this representation the one-column DKP wave function and its projections
are given by
\begin{align}
\psi &  =\left(
\varphi, 
\psi^{\mu}
\right)^{\intercal} ,\qquad\quad P\psi=\left(
\varphi, 
\mathbb{O}
\right)^{\intercal}  ,\qquad\quad P\tau\psi=\left(
\alpha\varphi, 
\mathbb{O}\right)^{\intercal}  ,\label{ap3a}\\
P^{\mu}\psi &  =\left(
\psi^{\mu}, 
\mathbb{O}
\right)^{\intercal}  ,\quad\qquad P^{\mu}\tau\psi=\left(
(1-\alpha)\psi^{\mu}, 
\mathbb{O}
\right)^{\intercal}  \label{ap3b}%
\end{align}
where $\mathbb{O}\equiv \left[  0\right]  _{4\times1}$. 
The condition $\tau^{2}-\tau=0$ implies for the $\alpha$ parameter that 
$
\alpha^{2}-\alpha=0\Rightarrow\alpha=0,1\,.$ The value $\alpha=1$ corresponds to a topological field, while $\alpha=0$
reproduces the massless Klein-Gordon field \cite{HCC}. In this representation the explicit relations among the
components of the massless spin $0$ DKP field are $
  (1-\alpha)\psi^{\mu}=i\partial^{\mu}\varphi\,$ and $\alpha\varphi=i\,\partial_{\mu}\psi^{\mu}$. For $\alpha=0$ the DKP
Lagrangian density (\ref{lag}) reduces to the usual one for the massless
Klein-Gordon field
$\mathcal{L}=\partial^{\mu}\varphi^{\ast}\partial_{\mu} \varphi\,$. For more aspects on the equivalence between DKP and the Klein-Gordon see, for instance, Refs. \cite{Fainberg:1999ca,Castro:2014lxa}. Moreover, an interesting approach of the Dirac and DKP equations for  spin 1/2 and spins 0 and 1, respectively, as the simplest special cases of the Bhabha system of first order relativistic wave equations for arbitrary spin can be found at  \cite{Krajcik:1976tv,Fischbach:1974cy}, and references therein. The DKP algebra has been approached furthermore in other interesting contexts \cite{Nedjadi:1993uq,Nedjadi:1993un}.

The DKP field can be related to the tensor product of two Dirac spinor fields, and endows their symmetries under  charge conjugation $\mathscr{C}$, parity $\mathscr{P}$, and time reversal $\mathscr{T}$.
Although in \cite{photon} the $A^\mu$ and the $F^{\mu\nu}$ are respectively interpreted as the electromagnetic potential and the field strength, the formalism is  thoroughly general.
It means that for the standard DKP field, the charge conjugation, parity, and time reversal are involutions, when acting on the scalar, the 4-vector,  and the antisymmetric tensor of the DKP field components in (\ref{psid0}, \ref{psid1}). In particular, $\mathscr{P}^2 = \mathbb{I} = (\mathscr{CPT})^2$.
 
 DKP fields given by Eqs. (\ref{psid0}) and (\ref{psid1}) can be expressed as the tensor product of two algebraic spinor fields, namely, elements of a minimal left ideal in the Clifford-Dirac algebra $\mathbb{C}\ell_{1,3}$. 
In order to realize the relationship between Clifford algebras and DKP algebras \cite{Jacobson, micali}, let us take a quadratic space $(V,g)$, which can be thought as being the tangent space at a point on a Lorentzian manifold. Let us also consider $\mathcal{A}$ an associative algebra with unity $1_{\sst \mathcal{A}}$ and let $\gamma$ be the 
 linear application $\gamma: V \rightarrow \mathcal{A}$. 
The pair 
$(\mathcal{A},\gamma)$ 
is a {Clifford algebra\/}\index{\'Algebra! de Clifford} $\cl(V,g
)$ for the quadratic space 
$(V,g)$ when $\mathcal{A}$ is generated as an algebra by $\{\gamma({}{v})
\,|\, {}{v}\in V\}$ and $\{a1_{\sst \mathcal{A}}\,|\, a \in \mathbb{R}\}$, satisfying $
{ \gamma({}{v})\gamma({}{u}) + 
\gamma({}{u})\gamma({}{v}) = 
2 g({}{v},{}{u})1_{\sst \mathcal{A}} }$, 
for all ${}{v},{}{u} \in V$. For a basis $\{e_\mu\}$ of $V$, the element $\gamma(e_\mu)$ is usually denoted by $\gamma_\mu$ \cite{lou2}. 

 Throughout this paper $V$ is considered to be the Minkowski spacetime $\mathbb{R}^{1,3}$. The Clifford-Dirac algebra for this particular case is denoted by $\mathbb{C}\ell_{1,3}$.
Spinors are well known to be elements of a minimal ideal of $\mathbb{C}\ell_{1,3}$ \cite{lou2}. The composed spinor describing an element of the DKP algebra is therefore an  element of  $\mathbb{C}\ell_{1,3}\otimes\mathbb{C}\ell_{1,3}$.  By considering the mapping $\delta: \mathbb{R}^{1,3} \rightarrow \mathbb{C}\ell_{1,3}\otimes\mathbb{C}\ell_{1,3}$ defined by \beq\label{deltav}\delta(v) = \frac{1}{2}(v\otimes1+1\otimes v)\eeq\noindent (here ``1" denotes the identity in $\mathbb{C}\ell_{1,3}$), the property
\beq
 \delta(u)\delta(v)\delta(u) = \frac{1}{8}(uvu\otimes1+u^2 \otimes v+(uv+vu)\otimes u+u\otimes(uv+vu)+v\otimes u^2 +1\otimes uvu)\eeq\noi can be derived \cite{micali}.
The Clifford relation
 $uv + vu = 2g(u,v)\,1$ implies $uvu =
2g(u,v)u-g(u,u)v$ and consequently  
$$
 \delta(u)\delta(v)\delta(u) = g(u,v)\delta(u).$$ By the universal property of the spacetime DKP algebra $B_{1,3}$, the application $\delta$ extends to an algebra monomorphism $\Delta: B_{1,3} \rightarrow \mathbb{C}\ell_{1,3} \otimes \mathbb{C}\ell_{1,3}$ \cite{Jacobson}, mapping every $\psi\in B_{1,3}$ in $\frac{1}{2}(\psi\otimes1+1\otimes\psi)$. It is worth mentioning that, for every $v\in \mathbb{R}^{1,3}$, $\Delta(2v^2-g(v,v)) = v\otimes v$. The DKP  algebra $B_{1,3}$ is thus the subalgebra of  $\mathbb{C}\ell_{1,3}\otimes\mathbb{C}\ell_{1,3}$ generated by all elements $\delta(v)$ \cite{Jacobson, micali}.
 
DKP fields are usually written as the tensor product of two Dirac spinor fields. Hereupon we shall depart from this assumption, and evince the unexpected role of introducing Elko spinor fields as the DKP fields building blocks. We prove that the interplay between Dirac and Elko spinor fields manifest VSR symmetries in the spin-0 and spin-1 DKP algebra sectors, with the aid of the graded tensor product. Firstly, let us brief revisit Elko spinor fields.

\section{Elko spinor fields}

In this Section some properties of Elko
spinor fields  are briefly
revisited. An Elko can be expressed in general as   \cite{allu,crevice}
\begin{equation}
\lambda({k^\mu})=\binom{i\Theta\phi^{\ast}(k^\mu)}{\phi(k^\mu)}, \qquad k^\mu \equiv \lim_{p\to 0}\left(m,{{\bf p}}\right),
\label{1}%
\end{equation}
\noindent  where $\phi(k^\mu)$ denotes a left-handed Weyl
spinor and $p = \|{\bf p}\|$. Given the rotation generators denoted by
${\mathscr{J}}$, the Wigner's spin-1/2 time reversal operator
$\Theta$ satisfies $\Theta
\mathscr{J}\Theta^{-1}=-\mathscr{J}^{\ast}$. Hereon, as in
\cite{allu}, the Weyl representation of $\gamma^{\mu}$
\begin{equation}
\gamma^{0}=%
\begin{pmatrix}
\OO_2 & \II_2\\
\II_2 & \OO_2
\end{pmatrix}
,\quad \gamma^{k}=%
\begin{pmatrix}
\OO_2 & -\sigma_{k}\\
\sigma_{k} & \OO_2
\end{pmatrix}\label{dirac matrices}
\end{equation}
\noindent  is used, where
 $\sigma_i$ are the Pauli matrices.    Elko spinor fields are eigenspinors of the charge
conjugation operator $C$, namely, $C\lambda(k^\mu)=\pm \lambda(k^\mu)$. 
The plus [minus] sign regards {self-conjugate} [{anti self-conjugate}]  spinor fields, denoted by $\lambda^{S}(k^\mu)$ [$\lambda^{A}(k^\mu)$]. Explicitly, the complete form of Elko spinor fields can be
found by solving the equation of helicity
$(\sigma\cdot\widehat{\bf{p}})\phi^{\pm}(k^\mu)=\pm
\phi^{\pm}(k^\mu)$ in the rest frame and subsequently
performing a boost \cite{allu, crevice}. 
Therefore Elko 
spinor fields are given by \cite{crevice}
\begin{align}
\lambda^{S}_{\pm}(p^\mu)&=\sqrt{\frac{E+m}{2m}}\Bigg(1\mp
\frac{{p}}{E+m}\Bigg)\lambda^{S}_{\pm}(k^\mu),
 &
\lambda^{A}_{\pm }(p^\mu)&=\sqrt{\frac{E+m}{2m}}\Bigg(1\pm
\frac{{p}}{E+m}\Bigg)\lambda^{A}_{\pm }(k^\mu),
\end{align}
where $\lambda^{S}_{\pm}(k^\mu)=%
\begin{pmatrix}
 i \Theta[\phi^{\pm}(k^\mu)]^{*} \\
\phi^{\pm}(k^\mu)
\end{pmatrix}$ and $\lambda^{A}_{\pm}(k^\mu)=%
\pm\begin{pmatrix}
-i \Theta[\phi^{\mp}(k^\mu)]^{*} \\
\phi^{\mp}(k^\mu)
\end{pmatrix}$. Moreover, the notation
  \begin{eqnarray}
\phi^+_L(k^\mu) && = \sqrt{m} \left(
									\begin{array}{c}
									\cos(\theta/2)\exp(- i \phi/2)\\
									\sin(\theta/2)\exp(+i \phi/2)
											\end{array}
									\right),\quad
\phi^-_L(k^\mu)  = \sqrt{m} \left(
									\begin{array}{c}
									-\sin(\theta/2)\exp(- i \phi/2)\\
									\cos(\theta/2)\exp(+i \phi/2)
											\end{array}
									\right) \nonumber
									\end{eqnarray} is used \cite{crevice}.

There are several interesting and unusual aspects concerning Elko theory. Most importantly for our purposes here is that the underlying Elko properties are ruled by VSR subgroups. Cohen and Glashow argued that Very Special Relativity, rather than Special Relativity, could be the fundamental symmetry of nature at the planck scale, with the standard model emerging as an effective theory \cite{cohen}. Ahluwalia and Horvath, however, proposed that VSR could appear at the standard model scale, related to dark matter \cite{horv1}. 

It is well known that when the parity operator is absent in a given relativistic theory, it is possible to rebuilt the dynamical objects by thinking of irreducible representations of  subgroups of the Lorentz group \cite{cohen}. The situation is the following: by removing the parity operator from the full Lorentz group (therefore working out of the discrete symmetries scope) it is possible to rearrange some of the boosts and rotations generators, giving rise to subgroups other than the orthochronous proper one. 
There are four possible groups, but we are mainly interested in the so called homotheties and similitude groups, denoted by HOM(2) and SIM(2) respectively.  

Explicitly, given $\mathcal{J}$ and $\mathcal{K}$ the rotation and boost generators, the generators $T_1:=K_x+J_y$ and $T_2:=K_y-J_x$ can be defined. Adding $K_z$ to those generators yields  the algebra $\mathfrak{hom}(2)$, which is a 3-parameter algebra associated with the
group of homotheties, HOM(2). On the other hand, adding $J_z$ to $\mathfrak{hom}(2)$ lead to a wider algebra associated to the similitude group, SIM(2). VSR is realized taking HOM(2) and SIM(2) as the symmetry groups of the theory. In fact, Cohen and Glashow have shown that the group HOM(2), which is a subgroup of SIM(2), is necessary and
sufficient to  \cite{cohen}: a) explain the results of the Michelson-Morley experiments and its more sensitive results; b) ensure that the speed of light is the same for all observers; c) preserve SR time dilatation and the law of velocity addition.
Furthermore the Elko theory (more precisely, the Elko spin sums) are invariant under the action of HOM(2) and covariant under SIM(2) \cite{horv1}.  

An enlightening review of VSR and why this symmetry must emerge in Elko theory can be found in the reference \cite{horv1}. For VSR in a wider class of Lorentz violation theory see \cite{rivelles}. 

In what follows, we shall investigate the typical signature of VSR subgroups in DKP algebra by using Elko spinor fields as building blocks of the DKP field. Notice, however, that in this case we must focus our attention naturally in the action of the charge conjugation operator on the spinor fields.

\section{Elko and Dirac spinor fields: the DKP algebra}\label{secdkp}

 This Section is devoted to introduce class-preserving Elko and Dirac spinor fields -- under Lounesto spinor field classification -- in order to express DKP fields in terms of the tensor product between Elko and Dirac spinor fields. It implies that Elko spinor fields can manifest different properties on the spin-0 and spin-1 sectors of DKP fields. Besides, the DKP algebra is shown to have an unexpected behavior,  
 when expressed as the graded  tensor product between Elko and Dirac spinor fields. 
 
Let us start by considering the Minkowski spacetime $M$ and $\{\mathbf{e}_{\mu }\}$ a section of the frame bundle $\mathbf{P}_{%
\mathrm{SO}_{1,3}^{e}}(M)$ and 
$\{\theta ^{\mu }\}$  be respectively its associated  dual basis. Classical spinor
fields carrying a $D^{(1/2,0)}\oplus D^{(0,1/2)}$ representation of SL$(2,%
\mathbb{C)\simeq }\;\,\mathrm{Spin}_{1,3}^{e}$ are sections of the vector
bundle $\mathbf{P}_{\mathrm{Spin}_{1,3}^{e}}(M)\times _{\rho }\mathbb{C}%
^{4}, $ where $\rho $ stands for the $D^{(1/2,0)}\oplus D^{(0,1/2)}$
representation of SL$(2,\mathbb{C) }$ in the complex $4\times 4$ matrices. Within this formalism, it is possible to make use of the multivector structure and write down a mother spinor field given by \begin{equation}\psi \sim (\sigma + {\bf J}+i{\bf S}-i\gamma_{0123}{\bf K}+\gamma_{0123}\omega)\eta,\end{equation} where $\eta$ is a spinor and $\sigma$, ${\bf J}$, ${\bf S}$, ${\bf K}$, and $\omega$ are the bilinear covariants provided by: 
\begin{align}
\sigma & =\psi ^{\dagger }\gamma _{0}\psi ,\quad \mathbf{J}=J_{\mu }\theta
^{\mu }=\psi ^{\dagger }\gamma _{0}\gamma _{\mu }\psi \theta ^{\mu },\quad 
\mathbf{S}=S_{\mu \nu }\theta ^{\mu \nu }=\frac{1}{2}\psi ^{\dagger }\gamma
_{0}i\gamma _{\mu \nu }\psi \theta ^{\mu }\wedge \theta ^{\nu },  \notag \\
\mathbf{K}& =K_{\mu }\theta ^{\mu }=\psi ^{\dagger }\gamma _{0}i\gamma
_{0123}\gamma _{\mu }\psi \theta ^{\mu },\quad \omega =-\psi ^{\dagger
}\gamma _{0}\gamma _{0123}\psi .  \label{fierz}
\end{align}%
Not every combination of bilinear covariants  generate different types of spinors, since the bilinear covariants satisfy the Fierz identities:
\begin{equation}
\mathbf{K}^{2}+\mathbf{J}%
^{2}=0=\mathbf{J}\cdot\mathbf{K},\qquad\mathbf{J}^{2}=\omega^{2}+\sigma^{2},\qquad\mathbf{K}\wedge\mathbf{J}%
=(\omega+\sigma\gamma_{0123})\mathbf{S}.  \label{fi}
\end{equation}
\noindent 
It turns out that the Lounesto spinor field classification provides just the following spinor field
classes \cite{lou2}, where in the first three classes clearly $\mathbf{J}, \mathbf{S}, \mathbf{K}\neq0$:
\begin{itemize}
\item[1)] $\sigma\neq0,\;\;\; \omega\neq0$\qquad\qquad\qquad\qquad\qquad4) $\sigma= 0 = \omega, \;\;\;\mathbf{K}\neq 0, \;\;\;\mathbf{S}\neq0$%
\label{Elko11}
\item[2)] $\sigma\neq0,\;\;\; \omega= 0$\label{dirac1}\qquad\qquad\qquad\qquad\qquad5) $\sigma= 0 = \omega, \;\;\;\mathbf{K}=0,\;\;\; \mathbf{S}\neq0$%
\label{tipo41}
\item[3)] $\sigma= 0, \;\;\;\omega\neq0$\label{dirac21} \qquad\qquad\qquad\qquad\qquad\!6) $\sigma= 0 = \omega, \;\;\; \mathbf{K}\neq0, \;\;\; \mathbf{S} = 0$%
\end{itemize}
\noi
Lounesto spinor field classes are well known not be preserved by sum of spinor fields. For instance, the sum of two Weyl spinor fields (type-(6)) are usually Dirac spinor fields (that  are encompassed by type-(1), -(2), and -(3)) under this classification.
We shall analyze the space generated by tensor product of Elko linear spaces and Dirac linear spaces. Thus, in what follows we aim to find the vector space structure underlying Elko and Dirac spinor fields spaces. This is necessary to guarantee that  the linear combination of spinor fields in the same class under Lounesto classification is still in the same spinor field class. In order to accomplish this for Elko spinor fields, we need to find a linear structure such that every spinor field must be either a type-(5) spinor field or the zero vector. Therefore an eigenspinor of the charge conjugation operator restricts to those corresponding to real eigenvalues.  A natural way to find this structure is considering the eigenspaces of the charge conjugation operator associated to positive or negative eigenvalues, namely, the self-conjugate class preserving and anti self-conjugate class preserving spaces, respectively. They are 4-dimensional real linear spaces:
\beq\mathcal{E}^S&=& {\rm span}_{\rm lin}^\mathbb{R}\,\{(-i, 0, 0, 1)^\intercal, (-1, 0, 0, i)^\intercal, (0, i, 1, 0)^\intercal, (0, 1, i, 0)^\intercal\}\,,\label{es}\\
\mathcal{E}^A&=& {\rm span}_{\rm lin}^\mathbb{R}\,\{(i, 0, 0, 1)^\intercal, (1, 0, 0, i)^\intercal, (0, -i, 1, 0)^\intercal, (0, -1, i, 0)^\intercal \}\,.\label{ea}\eeq\noi Explicit calculations can show that merely a few combinations mixing elements of both spaces are of type-(5), but they do not correspond to Elko.

For Dirac spinor fields (types-(1),-(2),-(3)), at least $\sigma$ or $\omega$ must be non null. Given two Dirac spinors 
\beq\label{dirf}\psi(x) = (\psi_1(x), \psi_2(x), \psi_3(x), \psi_4(x))^\intercal,\qquad\qquad\zeta(x) = (\zeta_1(x), \zeta_2(x), \zeta_3(x), \zeta_4(x))^\intercal\,,\nonumber\eeq\noi
where $\psi_\mu(x)$ and $\zeta_\mu(x)$ are complex functions, each one of the following conditions ensure that the sum of Dirac [Elko] spinors are Dirac [Elko] spinors.  In fact, for the sum of type-(1) Dirac spinors be type-(1) Dirac spinors, the conditions below must hold:
\beq
-\psi_1^*\psi_3-\psi_2^*\psi_4+\psi_3^*\psi_1-\psi_4^*\psi_2&\neq&-\zeta_1^*\zeta_3+\zeta_2^*\zeta_4+\zeta_3^*\zeta_1-\zeta_4^*\zeta_2\,,\label{23a}\\
\pm\psi_1^*\psi_3\pm\psi_2^*\psi_4+\psi_3^*\psi_1\pm\psi_4^*\psi_2&\neq&\zeta_1^*\zeta_3\pm\zeta_2^*\zeta_4+\zeta_3^*\zeta_1\pm\zeta_4^*\zeta_2\,,\label{23b}\\
\pm\psi_1^*\psi_3\pm\psi_2^*\psi_4\pm\psi_3^*\psi_1\pm\psi_4^*\psi_2&\neq&\zeta_1^*\zeta_3\pm\zeta_2^*\zeta_4+\zeta_3^*\zeta_1\pm\zeta_4^*\zeta_2\,,\label{23c}
\eeq
\beq
-\psi_1^*\psi_4\pm \psi_2^*\psi_3\mp\psi_3^*\psi_2+\psi_4^*\psi_1&\neq&-\zeta_1^*\zeta_4\pm\zeta_2^*\zeta_3\mp\zeta_3^*\zeta_2+\zeta_4^*\zeta_1\,,\label{31a}\\
\psi_1^*\psi_4\pm \psi_2^*\psi_3+\psi_3^*\psi_2\pm\psi_4^*\psi_1&\neq&\zeta_1^*\zeta_4\pm\zeta_2^*\zeta_3+\zeta_3^*\zeta_2\pm\zeta_4^*\zeta_1\,,\label{31b}
\eeq
\beq
\lVert\psi_1\lVert^2\pm \lVert\psi_2\lVert^2\mp\lVert\psi_2\lVert^2-\lVert\psi_4\lVert^2&\neq&\lVert\zeta_1\lVert^2\pm \lVert\zeta_2\lVert^2\mp\lVert\zeta_2\lVert^2-\lVert\zeta_4\lVert^2\,.\label{311}
\eeq
For type-(2) Dirac spinors, one of the conditions in (\ref{23a}, \ref{23b}, \ref{23c}), namely
\beq\psi_1^*\psi_3+\psi_2^*\psi_4+\psi_3^*\psi_1+\psi_4^*\psi_2&\neq&\zeta_1^*\zeta_3+\zeta_2^*\zeta_4+\zeta_3^*\zeta_1+\zeta_4^*\zeta_2,\nonumber\eeq can be relaxed, as well as  one of the conditions in (\ref{311})
\beq
\lVert\psi_1\lVert^2+ \lVert\psi_2\lVert^2-\lVert\psi_2\lVert^2-\lVert\psi_4\lVert^2&\neq&\lVert\zeta_1\lVert^2+ \lVert\zeta_2\lVert^2-\lVert\zeta_2\lVert^2-\lVert\zeta_4\lVert^2\,,\label{3113}\nonumber
\eeq\noi for type-(3) Dirac spinor fields. Actually, the conditions for the sum of the spinors $\psi(x)$ and $\zeta(x)$ to be a Dirac (regular) spinor can be summarized in the expression
\begin{equation}
(\psi_1+\zeta_1)( \psi_3+\zeta_3)^* \neq (\psi_2+\zeta_2)(\psi_4+\zeta_4)^*.
\end{equation}

These relations are also satisfied by two 4-dimensional complementary  linear spaces:
\beq 
\mathcal{D}^+&=& {\rm span}_{\rm lin}^\mathbb{R}\{(1, 
0, 
1,  0)^\intercal, (i, 0, i, 0)^\intercal, (0, i, 0, 1)^\intercal, (0, -1, 0, i)^\intercal\},\label{dp}\\
\mathcal{D}^-&=& {\rm span}_{\rm lin}^\mathbb{R}\{(-1, 0, 1, 0)^\intercal, (-i, 0, i, 0)^\intercal, (0, -i, 0, 1)^\intercal, (0, 1, 0, i)^\intercal\}\,.\label{dm}\eeq\noi These spaces contain all types of Dirac spinor fields. 
For Dirac fields the constraints $\mathbf{K} \neq 0$ or $\mathbf{S} \neq 0$ are not necessary:  Lounesto classification implies that there are no classes of spinors with $\sigma$ or
$\omega$ non null, and $\mathbf{K}$ or $\mathbf{S}$ null. In fact, it results from the implications $\mathbf{K}=0 \Rightarrow \sigma =0=
\omega$ and $\mathbf{S}=0 \Rightarrow \sigma =0=
\omega$. For an useful characterization of the spinor classes under Lounesto classification see \cite{spissue}, with a survey in \cite{unfo}. Such classification in spaces endowed with arbitrary bilinear forms is accomplished in \cite{abla}.


Hereon we shall denote by $\ddd$ [$\eee$] any of the spaces (\ref{dp}, \ref{dm}) [(\ref{es}, \ref{ea})]. Operators acting on the space $\mathcal{S}_1\ot \mathcal{S}_2$, where $\mathcal{S}_i \in \{ \ddd,\eee \}$, $i=1,2$, are constructed as follows:
\beq
{\mathscr{O}} = \frac{1}{\sqrt{2}}(\oo_{\sss _1}\ot \II_{\sss _2} + \II_{\sss _1}\ot\oo_{\sss_2})\,,\label{op1a}
\eeq\noi where $\oo_{\sss _i}\in$ End($\sss _i)$ denotes general operators in the endomorphism group acting on either the Dirac or Elko class preserving spinor fields space. We choose the above  factor $\frac{1}{\sqrt{2}}$  in order that the discrete symmetries act on the tensor product of spinor fields as involutions, for some cases. Hereupon our notation is devoid of the subindexes ${(\,\cdot\,)}_{{\sss}_i}\,$, ${(\,\cdot\,)}_{\ddd}\,$, or ${(\,\cdot\,)}_\eee$, which shall be implicitly noticeable.
 
Given $\chi_i\in\sss_i$, the successive action of the operator ${\mathscr{O}}\in$ End($\sss_1\ot\sss_2$)  respectively given by
\beq
\mathscr{O}(\chi_1\ot\chi_2)&=&\frac{1}{\sqrt{2}}(\oo\chi_1\ot\chi_2 + \chi_1\ot\oo\chi_2\label{o1}),\\
\mathscr{O}^2(\chi_1\ot\chi_2) 
&=& \frac{1}{2}\oo^2\chi_1\ot\chi_2 + \oo\chi_1\ot\oo\chi_2+\frac{1}{2}\chi_1\ot\oo^2\chi_2\,.\label{o2}\eeq\noi
In general it is straightforward to show that 
\beq
\mathscr{O}^n(\chi_1\ot\chi_2)=2^{-\frac{n}{2}}\sum_{p=0}^{n}{n \choose p} \oo^{n-p}\chi_1\ot\oo^p\chi_2.\nonumber
\eeq
The commutator and the anti-commutator 
of two operators $\mathscr{O}_i=\frac{1}{\sqrt{2}}(\oo_i\ot\II+\II\ot\oo_i)$,  acting on the  space $\sss_1 \ot \sss_2$ are thus provided by
\beq
[\mathscr{O}_1,\mathscr{O}_2]&=&\frac{1}{2}([\oo_1,\oo_2]\ot\II+\II\ot[\oo_1,\oo_2])\,,\\
\{\mathscr{O}_1,\mathscr{O}_2\}&=&\frac{1}{2}(\{\oo_1,\oo_2\}\ot\II+2\oo_1 \ot\oo_2+2\oo_2 \ot\oo_1+\II \ot\{\oo_1,\oo_2\})\,.
\eeq

As Dirac and Elko spinor fields 
are elements of  minimal left ideals in the Clifford-Dirac algebra $\mathbb{C}\ell_{1,3}$, they are not {a priori} even elements,  under the graded involution that defines the $\ZZ_2$-grading of $\mathbb{C}\ell_{1,3}$. Therefore, the graded tensor product can be defined. 
Actually, in the four dimensional space-time the map given in the equation (12) does not define a morphism when we change the standard tensor product to the graded one. When the domain is restricted, we it does define a morphism, in particular in the spinor operator context. 

%
%

 The {alternating tensor product} 
$\mathbb{C}\ell_{1,3}\,\hat{\otimes}\,\mathbb{C}\ell_{1,3}$ is the algebra generated by the product $a\,\hat{\otimes}\,b$, 
defined by
\begin{equation}
\label{eq.4.1}
(a_1\hat{\otimes}\, b_1) (a_2\hat{\otimes}\, b_2) = 
(-1)^{\deg{(b_1)}\deg{(a_2)}} a_1 a_2\,\hat{\otimes}\, b_1 b_2,\quad\quad a_1, a_2, b_1, b_2\in \mathbb{C}\ell_{1,3}\,.
\end{equation}
Hence the analogue of Eq.(\ref{o1})
in this context provided by the graded tensor product reads 
\beq 
\mathscr{O}(\chi_1 \hat{\ot} \chi_2)&=& \frac{1}{\sqrt{2}}[(\oo\hat{\ot}\II)(\chi_1 \hat{\ot} \chi_2) + (\II\hat{\ot}\oo)(\chi_1 \hat{\ot} \chi_2)]\n
&=& \frac{1}{\sqrt{2}}[\oo\chi_1 \hat{\ot} \chi_2 + \chi_1 \hat{\ot}\, \oo \chi_2].
\eeq\noi This case and the former one presented in (\ref{o1}) are similar. Notwithstanding, it does not hold in general, when higher order compositions of $\mathscr{O}$ are regarded. Indeed
\beq
\mathscr{O}^2(\chi_1 \hat{\ot} \chi_2) &=& \frac{1}{\sqrt{2}}\mathscr{O}(\oo\chi_1 \hat{\ot} \chi_2 + \chi_1 \hat{\ot}\, \oo \chi_2) \n
&=& \frac{1}{{2}}[\oo^2\chi_1\hat{\ot} \chi_2 + \oo\chi_1\hat{\ot}\oo\chi_2+(-1)^{\deg{\oo}}\oo\chi_1\hat{\ot}\oo\chi_2
+\chi_1\hat{\ot}\oo^2\chi_2].\label{oo3}\eeq\noi

The parity, charge conjugation, and 
time reversion operators are respectively defined  as \cite{allu}
\beq
\pp=e^{i\phi}\gamma^0 \mathcal{R},\qquad \cc = i\gamma^2 \mathcal{K}, \qquad\mot = i\gamma^1\gamma^3 \cc,\eeq
\noi where in spherical coordinates the action of the operator $\mathcal{R}$ is $\{\theta\mapsto \pi-\theta, \phi\mapsto \phi+\pi, r\mapsto r\}$. The operator $\mathcal{K}$ is the complex conjugation  operator. At this point it is worth to call attention to the fact that although spinors  live in $\mathbb{C}\ell_{1,3}$, the operators are defined on the representation space and need to be written in terms of $\mathbb{C}\ell_{1,3}$. Here, solely $\mathcal{K}$ need to be adapted, by acting on an algebraic spinor $\varphi$   as $\mathcal{K} \varphi = \gamma^{013}\varphi^* (\gamma^{013})^{-1}$ \cite{lou2}. Consequently, $\cc, \pp$,  and $\mot   $ are all odd operators under the graded involution, and therefore
\begin{eqnarray}\label{ac2}
2\mathscr{C}^2&=&\cc^2 \hat{\ot} \II+\II \hat{\ot} \cc^2,\\ \label{ap2}
2\mathscr{P}^2&=&\pp^2 \hat{\ot} \II+\II \hat{\ot} \pp^2,\\\label{at2}
2\mathscr{T}^2&=&\mot^2 \hat{\ot} \II+\II \hat{\ot} \mot^2,
\end{eqnarray}
hold for the alternating tensor product. 

In order to analyze the behavior of class preserving spaces above obtained   under actions of combinations of $\mathscr{C},\mathscr{P},\mathscr{T}$ operators, we will explicitly compute, in the next sections, the action of these operators on all combinations for the tensor product between Dirac spinors. As Elko spinor fields manifest VSR symmetries, solely the action of the operator ${\mathcal{C}}$ is taken into account in Subsections IV.B and IV.C all possible actions arising upon such a possibility.

\subsection{The tensor product between Dirac spinor fields}

Starting with the parity operator $\mpp\in$ End($\ddd\ot\ddd$), Eq.(\ref{o2})
implies \beq 
\mpp^2(\dod) &=& \frac{1}{{2}}(\pp^2\dod + 2\pp\psi_1\ot\pp\psi_2+\psi_1\ot\pp^2\psi_2)\n
&=& \dod + \pp \psi_1 \ot \pp \psi_2 \label{p2},\eeq\noi since the parity operator acting twice on a Dirac spinor is the identity operator, i. e., $\pp_\ddd^2 = \II_\ddd$. 
Therefore, 
for all $n=1,2,\ldots$,  there is a periodicity mod 2 for the operator $\mpp\in$ End($\ddd\ot\ddd$), given by
\beq
\mpp^{2n-1} &=& 2^{n-3/2}(\pp\ot \II + \II\ot \pp),\qquad\qquad 
\mpp^{2n} = 2^{n-1}(\II\ot\II + \pp\ot\pp)\,.\label{38}
\eeq\noi
Moreover, 
the anti-commutators and commutators of the charge conjugation and the parity operators in the DKP algebra  are respectively given by
\beq
\{\mcc,\mpp\}&=&\cc\ot\pp + \pp\ot\cc\,,\nonumber\\ 
 {}[\mcc,\mpp]&=& \frac{1}{2}([\cc,\pp]\ot \II+\II\ot[\cc,\pp])\,.
\eeq
Analogously, when one takes the time reversal operator the following relations are obtained:
\beq
\{\mcc,\mttt\}&=&\cc\ot\ttt + \ttt\ot\cc\,,\nonumber\\ 
 {}[\mcc,\mttt]&=& \frac{1}{2}([\cc,\ttt]\ot \II+\II\ot[\cc,\ttt])\,.\label{time}
\eeq
Besides, as the Dirac spinor field is not an eigenspinor of the charge conjugation operator 
let us analyze for instance in what aspect the DKP field generated by two Dirac spinor fields 
is different of a DKP field constructed by the tensor product between Dirac and Elko spinor fields. For instance, the following expressions shall be useful in the next Subsections in order to realize the signatures of Elko in the DKP fields:
\beq\label{cp4}
(\cc\ot\pp)(\psi_1\ot\psi_2)&=&\cc\psi_1\ot\pp\psi_2\,,\\
(\cc\ot\cc)(\psi_1\ot\psi_2)&=&\cc\psi_1\ot\cc\psi_2\,.\label{cpp4}
\eeq\noindent

For the alternating tensor product, according to Eqs.  (\ref{ac2}-\ref{at2})
the analogue of Eq.(\ref{o2}) is given by
\beq
\mathscr{C}^2=\mpp^2=\mathscr{T}^2&=&\,\II \hat{\ot} \II\,.\label{bbk}\eeq\noi 
Hereon we shall show that this {is not} the case when at least one of the Dirac spinor fields in the tensor product, constituting a DKP field, is substituted by an Elko spinor field, as only the charge conjugate operator must be taken into account.

\subsection{The tensor product between Elko and Dirac spinor fields}

For the charge conjugation operator $\mathscr{C}\in$ End($\ddd\ot\eee$) defined by (\ref{op1a}) it follows that
\beq 
\mathscr{C}^2(\doe) &=& \frac{1}{2}(\mathcal{C}^2\doe + 2\mathcal{C}\psi\ot\mathcal{C}\lam+\psi\ot\mathcal{C}^2\lam)\n
&=& \mathcal{C}\psi\ot\mathcal{C}\lam+\doe\nonumber\\
&=& \pm\mathcal{C}\psi\ot\lam+\doe\label{p2},\eeq\noi since $\cc_\ddd^2 = \II_\ddd\,$ and $\cc_\eee^2 = \II_\eee$. The sign in the above equation regards $\mathcal{C}\lambda^S = +\lambda^S$ and $\mathcal{C}\lambda^A = -\lambda^A$. 

When we take into account a DKP field 
$\lambda\otimes\psi$, the action of the operator $(\cc\ot\pp)$ is given by
\beq\label{cp51}
(\cc\ot\pp)(\lam^S\ot\psi)&=&+\lam^S\ot\pp\psi\,,\\
(\cc\ot\pp)(\lam^A\ot\psi)&=&-\lam^A\ot\pp\psi\label{majj1}\,.
\eeq\noindent Comparing Eq.(\ref{cp4}) to this case, it can be seen that 
Eq.(\ref{majj1}) is a signature in the DKP field  inherent to the anti self-conjugate Elko, as Eq.(\ref{cp51}) provides a signature for the DKP field intrinsic to the Majorana field.  

In addition, when the action of the operator $\mathcal{C}\otimes\mathcal{C}$ on such type of DKP fields are the following
\beq\label{cp5}
(\cc\ot\cc)(\lam^S\ot\psi)&=&+\lam^S\ot\cc\psi\,,\\
(\cc\ot\cc)(\lam^A\ot\psi)&=&-\lam^A\ot\cc\psi\label{majj}\,.
\eeq\noindent

As the previous case, exactly the same relations provided by Eqs.(\ref{time}) hold if one substitutes the parity operator for the time reversal operator.
Indeed, by taking into account the alternating tensor product we have 
\begin{eqnarray}
\mathscr{C}^2 &=& \,\II\, \hat{\ot}\, \II.\label{umum}
\end{eqnarray}
Elko leaves a signature on the action of discrete symmetries on DKP fields that neither Dirac spinors nor Majorana ones can carry, as we can see for instance from Eq.(\ref{majj}). These results evince that the DKP field constructed by a tensor product between Dirac and Elko spinor fields is not invariant under the full Lorentz group. It shows how the Elko field induces the VSR symmetry on the DKP field. A similar case is analyzed in the next Subsection.

\subsection{The tensor product between Elko spinor fields}

The DKP algebra generated by the alternate tensor product of two Elko reflects the same properties of discrete symmetries acting on Elko spinor fields. 
As $\mathcal{C}\lambda^S = +\lambda^S$ and $\mathcal{C}\lambda^A = -\lambda^A$ the only possibility is to act such operator on Elko, to test the signature of Elko spinor fields on the DKP field
\beq
(\cc\otimes\cc)(\lam^S\ot\lam^S)&=&\lam^S\ot\lam^S\nonumber\\
(\cc\otimes\cc)(\lam^A\ot\lam^S)&=&-\lam^A\ot\lam^S\nonumber\\
(\cc\otimes\cc)(\lam^S\ot\lam^A)&=&-\lam^S\ot\lam^A\nonumber\\
(\cc\otimes\cc)(\lam^A\ot\lam^A)&=&\lam^A\ot\lam^A\label{ccaa}
\eeq
and therefore the self or anti-selfconjugacy of the Elko spinor fields further influence in the action of the operator $(\cc\otimes\cc)$ on the DKP field, leading to a typical trace.

\section{Final remarks: VSR structure in the DKP fields}
\label{aadkp}

Clifford algebras play a prominent role on the construction of DKP fields. Dirac spinor fields are regular under Lounesto spinor field classification, and they do not bring any novelty  about the subgroups of the Lorentz group, in the construction of the DKP fields. On the another hand, Elko spinor fields are used to incorporate the VSR symmetries, departing from the Lorentz paradigm. 
Cohen and Glashow have shown that time dilation, the law of velocity addition, and the universal and isotropic velocity, do not demand the entire Lorentz group. Instead, those properties can be evinced by VSR subgroups \cite{cohen}.  If any  of the discrete symmetries of $\cal{P, T, CP},$ or $\cal{CT}$ is violated, the symmetry group
 is isomorphic to VSR subgroups and the largest one  is obtained by adjoining the four spacetime translation generators to the 4-parameter subgroup SIM(2). 
 
We proved that  $\cal{C}, P$, and $\cal{T}$ symmetries are induced on DKP fields, from Dirac spinor fields, by Eq.(\ref{bbk}), naturally.  In this way 
the correct properties of DKP fields can be highlighted. Some similar behavior occurs for Elko spinor fields. 
The DKP fields in the spaces $\ddd\hat{\ot}\eee$ and $\eee\hat{\ot}\ddd$ induce the charge conjugation operator to be an involution (Eq.(\ref{umum})). 
    

As Elko spinors are ruled by VSR HOM(2) and SIM(2) symmetries, and as 
the DKP algebra is a subalgebra in $\mathbb{C}\ell_{1,3}$, the results obtained here reflect exactly that the parity symmetry is  broken, and then the standard DKP algebra is not recovered. It indicates a manifestation 
of the Elko symmetries on the spin-0 and spin-1 bosonic sectors of DKP algebra.
DKP fields can be constructed, in this structure, by the tensor product of spinor fields. When Elko spinors are taken into account in this construction, DKP fields are shown to manifest a different signature, under the action of operators constructed from the tensor product between charge conjugation operator (in every sector of the DKP resulting field) and the parity operator (concerning the Dirac part of the DKP field). Moreover, it would be interesting to investigate this connection in other dimensions, in particular, in the plane as well as in higher dimensions \cite{corson,nikitin,janieto}.

The main idea here studied was that since $\psi\otimes\lambda$ and $\lambda\otimes\lambda'$ have different discrete symmetries
to $\psi \otimes \psi'$, the vector fields associated with the former will be physically distinct from the
later. In a broader sense, the mathematization previously exposed is an attempt to explore the well known richness of couplings within the DKP scope \cite{Ant} in the light of VSR subgroups. In this context, the appearance and formalization of Elko spinors as a building block of the DKP resulting field is particularly suggestive.

We finalize by stressing that it is also interesting to consider the tensor product of the type $\Theta\phi^*\otimes \phi$, as denoted in Section III and show that this field have a well-defined kinematics that is different from the usual Lorentz-invariant vector field, what shall be addressed in a future publication.

 
\section*{Acknowledgments}
R. da Rocha is grateful to SISSA and Prof. Loriano Bonora for the hospitality, to CNPq grants 303027/2012-6 and 473326/2013-2, and is also \emph{Bolsista da CAPES Proc. n$^{o}$} 10942/13-0.
 J. M. Hoff da Silva thanks to CNPq (482043/2011-3; 308629/2012-6).
 R. T. Cavalcanti thanks to UFABC and CAPES for financial support.

\end{document}